\documentclass[prl, reprint, superscriptaddress, showpacs]{revtex4-2}

\usepackage{graphicx}
\usepackage{dcolumn}
\usepackage{bm}

\usepackage{amsmath}
\usepackage{amssymb}
\usepackage{amsfonts}
\usepackage{mathtools}
\usepackage{lastpage}
\usepackage[usenames,dvipsnames]{color}
\usepackage{epsf}
\usepackage{multirow}
\usepackage{url}
\usepackage{natbib}
\usepackage{setspace}
\usepackage{enumerate}
\usepackage{epstopdf}
\usepackage{wrapfig}
\usepackage{morefloats}
\usepackage{float}
\usepackage{braket}
\usepackage{slashed}

\usepackage[T1]{fontenc}			
\usepackage[utf8]{inputenc}			
\usepackage{cancel}                             		

\begin{document}

\title{
Behavior of kinetic instabilities in a dynamically forming resonant distribution}

\author{E. J. Hartigan-O'Connor}
\affiliation{%
 Princeton Plasma Physics Laboratory, Princeton University, Princeton, NJ, 08543, USA}%

\author{T. Barberis}
\affiliation{%
 Princeton Plasma Physics Laboratory, Princeton University, Princeton, NJ, 08543, USA}%

\author{E. G. Devin}
\affiliation{%
 Princeton Plasma Physics Laboratory, Princeton University, Princeton, NJ, 08543, USA}%

\author{A. Bierwage}
\affiliation{%
 National Institutes for Quantum Science and Technology (QST), Naka Institute for Fusion Science and Technology, Ibaraki 311-0193, Japan}%
 
 \author{V. N. Duarte}%
 \email{vduarte@pppl.gov}
\affiliation{%
 Princeton Plasma Physics Laboratory, Princeton University, Princeton, NJ, 08543, USA}%

\date{\today}

\begin{abstract}
Instabilities driven by energetic particles are central to the physics of a burning plasma. The majority of kinetic simulations and reduced models assume that the unstable distribution is already fully established when energetic-particle-driven modes grow unstable. In realistic scenarios, however, energetic particles may accumulate in the resonance on an effective timescale comparable to the growth rate of the instability, meaning that the formation of the resonant distribution and the growth of the unstable mode must be treated concurrently. We study the behavior of these instabilities in the presence of such a dynamically forming distribution, evaluating two distinct metrics which measure how close a mode is to its linear stability threshold and how close a mode remains to its nonlinear stability threshold. It is found that saturation at large $\omega_b/\nu_\text{eff}$ (where $\omega_b$ is the bounce frequency of deeply trapped particles and $\nu_\text{eff}$ is the effective scattering rate at a resonance), normally associated with strongly driven excitation, can be achieved even if dynamically the mode remains at all times near its nonlinear stability threshold. We extend existing analytic models for near-marginal and far from marginal modes allowing for a time-dependent linear growth rate, deriving explicit expressions for the mode amplitude evolution. These formulas are shown to agree with nonlinear kinetic simulations.  The discrepancies between the case of a dynamically forming distribution and the case of a fully formed distribution are shown to be particularly pronounced for energetic particle distributions which relax diffusively.

\end{abstract}

\maketitle


\textit{Introduction}\textemdash Nonlinear kinetic instabilities, which arise via phase-space resonances with non-equilibrium features of an energetic population distribution, underpin a myriad of transport phenomena in plasmas \cite{LandauLifshitzVol10, Davidson1972, Melrose1986, Swanson2003}. In fusion experiments, they can be driven by gradients produced by beam injection, radiofrequency heating and fusion-born alpha particles, and are known to play a defining role in the predictability of plasma performance \cite{Heidbrink2008, ChenZoncaRevModPhys.2016, Salewski2025Energetic}. 

The sustainment of burning plasmas relies on these energetic particles (EPs) being well confined, to enable them to transfer the bulk of their energy to the reacting thermal species \cite{fasoli2007physics}. EPs can, however, resonantly excite collective modes to large amplitudes, leading to enhanced losses. The resulting confinement degradation depends on the saturation levels of the instability, which is determined by the balance between sources driving the instability and wave damping on the thermal plasma background \cite{breizman_sharapov_2025}. 

In experiments, the global EP population is formed dynamically due to the presence of a source, including external heating systems, and reaches a steady state on collisional timescales \cite{Gaffey1976JPP}. The wave-particle resonance description has largely employed the assumption that the underlying resonant distribution is already fully formed when the instability starts to develop (e.g., \cite{Landau1946,ONiel1965,Mazitov1965,BerkPRL1996,BerkPLA1997}). The accumulation of particles near narrow resonances, however, can occur on effective timescales comparable to the instability growth itself \cite{WongBerkSaturationinTFTRPoP1997, FasoliPRL1998, Gorelenkov1999Saturation, DuarteAxivPRL}, meaning that the unstable distribution is still forming while the mode grows \footnote{Interestingly, plasmoid instabilities in forming current sheets have been studied, showing that the instability onset prevents the achievement of a fully formed distribution \cite{Comisso_2017}. This can be seen as a fluid analogy of this present kinetic study.}. In this Letter, we address this gap by formulating a framework for nonlinear kinetic instabilities evolving concurrently with the formation of their own drive, revealing how time-dependent distribution buildup reshapes both the pathway to saturation and their apparent proximity to marginal stability. 



A slow formation of the resonant distribution has been invoked as a justification for instabilities to remain near marginal stability \cite{berk_2012_aip, breizman_sharapov_2025}. This assumption allows for considerable simplification as the kinetic equation can be expanded in powers of the mode amplitude \cite{BerkPRL1996,BerkPPR1997}. Despite the assumed  slowly forming distribution, studies of near-marginal kinetic instabilities have employed a fully formed distribution at the outset. Further work is therefore needed to assess the consistency of the marginality assumption and to justify its application in analytical models when the formation of the distribution and the growth of the instability are simultaneous.

Our results demonstrate that quasi-steady saturation in the nonlinear phase can mask strongly driven behavior as near-threshold evolution, while preserving the central role of the ratio between nonlinear trapping and effective scattering in categorizing the excitation regime. This extends the applicability of reduced kinetic models to more realistic scenarios with a time-dependent underlying distribution. The results reveal that the background noise level in comparison to the fueling rate of resonant particles is a critical parameter in determining the instability evolution character. This Letter also reports compact analytic formulas for the amplitude evolution in a forming distribution and discusses their consequences to fusion experiments.





\textit{Nonlinear simulations}\textemdash 
We use a one-dimensional kinetic model to investigate the behavior of kinetic instabilities concurrent with a forming resonant distribution. The instantaneous linear growth rate is given by $\gamma_L(t) \equiv 2\pi^2 (e^2\omega/mk^2) \partial F_0(t,v_{\text{res}})/\partial v$, where $F_0(t,v)$ is the underlying forming resonant distribution in the absence of any mode-induced flattening, $\omega$ and $k$ are the mode frequency and wave number, $v$ is the velocity coordinate, and $v_{\text{res}}=\omega/k$ is the location of the resonance in phase space. To allow for different excitation regimes, $F_0(t,v)$ is initialized with zero gradient around the resonance and, in the absence of modes, will asymptotically reach an equilibrium \textit{target} distribution $F_T$ which will ultimately provide a linear growth rate $\gamma_T = 2\pi^2 (e^2\omega/mk^2) \partial F_T(t,v_{\text{res}})/\partial v$.  

We now introduce the dimensionless variables $\Omega \equiv (kv-\omega)/\gamma_T$ and $\xi \equiv kx-\omega t$, and define $\omega_b^2 \equiv ekE/m$ as the bounce frequency of the particles most deeply trapped in the wave, where $E$ is the wave electric field amplitude. Then, the kinetic equation becomes
\begin{equation} \label{eq:kinetic_eq}
\frac{1}{\gamma_T} \frac{\partial f}{\partial t} + \Omega \frac{\partial f}{\partial \xi} + A\text{cos}\xi \frac{\partial f}{\partial \Omega}= \frac{\nu_{\text{eff}}^3}{\gamma_T^3}\frac{\partial^2}{\partial \Omega^2} \left(f-F_T\right),
\end{equation} 
where the normalized mode amplitude is defined as $A \equiv \omega_b^2/\gamma_T^2$ and $\nu_\text{eff}$ \cite{BerkPPR1997, DuartePoP2017} represents the characteristic timescale on which particles scatter across a narrow resonance. The system additionally satisfies the power balance equation \cite{BerkPoP1996,BerkPLA1997,BerkPPR1997,PetviashviliThesis}:
\begin{equation} \label{eq:power_balance}
\frac{dA}{dt} + \gamma_d A = -\frac{2e^2 \omega}{mk\gamma_T} \int_{-\pi}^{\pi}d\xi e^{-i\xi} \int d\Omega f(t,\xi,\Omega).
\end{equation}
We simulate the mode dynamics by simultaneously solving Eqs. \ref{eq:kinetic_eq} and \ref{eq:power_balance} using the BOT code \cite{Lilley2010}, which we modified to allow for an initial condition distinct from its standard implementation (which assumes a distribution with a constant slope in $\Omega$-space). To do so, we take a target distribution $F_T = mk\gamma_T^2(\omega+\Omega)/2\pi^2 e^2 \omega$ and explicitly solve the kinetic equation (Eq. \ref{eq:kinetic_eq}) in the absence of any modes, for a collision operator containing only diffusive scattering: $\partial F_0/\partial t = (\nu_{\text{eff}}^3/\gamma_T^2)\partial^2 \left(F_0-F_T\right)/\partial \Omega^2.$ The initial condition is taken as a zero gradient in the vicinity of the resonance, which is then collisionally relaxed towards the linear target distribution. (see Appendix A for details on these modifications). We find that the linear growth rate near the resonance is given by: 
\begin{equation} \label{eq:gammaLt}
\gamma_L(t) =\gamma_T\left[1-\alpha^{3/2}/\left(\nu_{\text{eff}}^3t/{\gamma_T^2} +\alpha \right)^{3/2}\right],
\end{equation}
where the dimensionless parameter $\alpha$ represents the scale of inhomogeneity of the distribution. Physically, $\sqrt{\alpha}$ corresponds to the distance in $\Omega$-space between the resonance and the initial EP source for a physical model which includes the full distribution outside the resonance. We can also interpret $\alpha$ as proportional to the normalized timescale for $\gamma_L$ to approach $\gamma_T$. For $\alpha \gg 1$, the relaxation is very slow, corresponding to the slow buildup of the EP distribution around the resonance when the resonance is far from the injected EPs. In the opposite limit, the distribution reaches its target state on a shorter timescale, as if the resonance is very close to the EP source. For this study, we limit our focus to modes with steady saturation levels, not focusing on strongly nonlinear cases like chirping \footnote{By keeping the collision rate much lower than the mode growth rate corresponding to the distribution, Ref. \cite{Vann2007} examined the bursty response of kinetic instabilities in a forming distribution, which precluded saturation. It was found that the wave frequency chirping maintains an otherwise strongly driven distribution near marginal stability.}.

To interpret the results, we also define two figures of merit for estimating how close modes are to threshold over time. We define 
\begin{equation} \label{eq:marg_params}
M_L(t) \equiv \frac{\gamma_L(t) - \gamma_d}{\gamma_L(t)}~,~M_{NL}(t) \equiv \frac{\gamma_{NL}(t) - \gamma_d}{\gamma_L(t)},
\end{equation}
where the nonlinear growth rate $\gamma_{NL}(t)$ is defined via $dA(t)/dt=(\gamma_{NL}(t)-\gamma_d)A(t)$. $M_L(t)$ is the time-dependent version of the conventional marginal stability parameter $(\gamma_L - \gamma_d)/\gamma_L$ \cite{BerkPRL1996}. When $M_L(t) \ll 1$, the mode is near-marginal in the linear sense; when $M_L(t)$ approaches unity, the mode is strongly driven. The second quantity, $M_{NL}(t)$, measures how close the mode is to its instantaneous saturation level, and corresponds to the relative rate of energy conversion between the resonant distribution and the mode. $M_{NL}(t)$ and the total net growth rate $\gamma_{NL}(t) - \gamma_d$ can be inferred directly from measurements. 



Saturation levels have been derived for near \cite{BerkPPR1997,BreizmanPoP1997} and far \cite{PetviashviliThesis,BerkBreizman1990a,devin2025emergence} from marginal cases where $\gamma_L$ is constant. Here, we allow $\gamma_L$ to be time-dependent, to analyze whether modes are near a quasi-steady saturation level during their dynamical evolution:
\begin{equation} \label{eq:A_near}
A_{\text{near}}(t) \equiv 1.39\left(\nu_\text{eff}^2/\gamma_T^2\right)\left( 1-\gamma_d/\gamma_L(t)\right)^{1/2},\end{equation}
\begin{equation}\label{eq:A_far}
A_{\text{far}}(t) \equiv 1.44\left(\nu_\text{eff}^2/\gamma_T^2\right)\left(\gamma_L(t)/\gamma_d\right)^{2/3}.
\end{equation}

\begin{figure}[ht]
\centering
\includegraphics[width=\columnwidth]{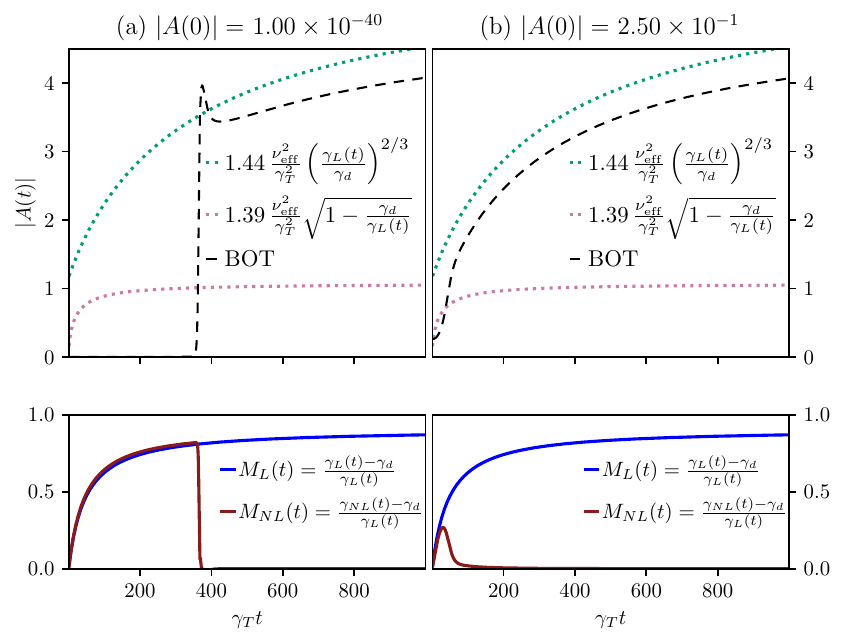}
\caption{BOT run for (a) small and (b) large initial amplitude $\omega_b(0)/\gamma_T$. Plotted are the instantaneous linear growth rate $\gamma_L(t)$, near-saturation parameter $M_{NL}(t)$, and mode amplitude. Time-dependent expected saturation levels for near-marginal and strongly driven modes are plotted in pink and green respectively.}
\label{fig:BOT_run_combined}
\end{figure}

Fig. \ref{fig:BOT_run_combined} studies the impact of the initial mode amplitude on the subsequent dynamics, for otherwise identical setups which eventually reach the strongly driven regime. Fig. \ref{fig:BOT_run_combined}(a) shows the results of a BOT simulation for a mode with $\gamma_d/\gamma_T = 0.1$, $\nu_{\text{eff}}/\gamma_T = 0.9$, $\omega_b(0) \sim 10^{-20} \gamma_T$, $(1/\gamma_T)d\gamma_L(t=0)/dt \sim 10^{-4} \gamma_T$, and $\alpha=500$. The initial mode amplitude is very small ($\omega_b(0) = 10^{-20}\gamma_T$), so the phase mixing is slow and the distribution builds up without interference from the mode until $t \gamma_T \approx 350$, when the mode grows linearly towards the strongly-driven saturation level. During this stage of linear growth, $M_{NL}(t)$ becomes closer to 1, meaning that the mode is far from quasi-steady saturation. After the linear growth, the mode tracks the strongly-driven saturation level (shown in green), which slowly increases with $\gamma_L(t)$ while $M_{NL}(t)$ remains near zero. 

Fig. \ref{fig:BOT_run_combined}(b) uses the same parameters, except that the mode is initialized close to its saturation amplitude ($\omega_b(0) = 0.5\gamma_T$). In this case, the distribution has much less time to build a steep gradient, and the linear growth phase of the mode evolution concludes when $M_{NL}(t)$ is still low. The quasi-steady saturation level the mode initially attains is relatively close to that of a marginally stable mode (shown in pink). As the background distribution continues to build up, the mode remains nonlinearly saturated ($M_{NL}(t) \ll 1$), but the amplitude slowly increases towards the strongly-driven saturation level. The long-term behavior of the modes in Fig. \ref{fig:BOT_run_combined}(a) and \ref{fig:BOT_run_combined}(b) is the same despite their different time histories, an observation that holds for a wide range of $\gamma_d/\gamma_T$, $\nu_{\text{eff}}/\gamma_T$, and $\omega_b(0)/\gamma_T$. This implies that modes may saturate at the far from linear threshold ($\gamma_T\gg\gamma_d$) level \cite{PetviashviliThesis,BerkBreizman1990a} while never departing significantly from its instantaneous nonlinear saturation ($\gamma_{NL}\approx\gamma_d$).

These results show that the saturation level of a mode is unaffected by its time history; however, the dynamical evolution of the mode may vary dramatically depending on the initial condition and the rate of formation of the resonant distribution. In the next sections, we focus on the characteristics of this transient behavior, and derive explicit analytic expressions for $A(t)$ which agree remarkably well with simulation results.


\textit{Analytic evolution of strongly driven modes}\textemdash With the growth rate defined in Eq. \ref{eq:gammaLt}, we derive analytic expressions for mode amplitude in two opposite limits. The first is the (linearly) strongly-driven limit, where $M_L(t) \approx 1$. We follow \cite{devin2025emergence} to divide the mode evolution into two phases, under a time-local approximation. Phase I describes the linear growth of the mode, with the mode amplitude given by:
\begin{equation} \label{eq:strong_solution_phase1}
A(t) = A(0) \text{exp}\left( \int_0^{t} dt'(\gamma_L(t')-\gamma_d)\right).
\end{equation}
Phase II describes the mode dynamics on a longer timescale when $\omega_b^3/\nu_{\text{eff}}^3 \gg 1$, after the mode grows to a sufficiently large amplitude in the linear phase. To first order in the small parameter $\nu_{\text{eff}}^3/\omega_b^3$, the amplitude is given by $d|A|/dt+\gamma_{d}|A|=1.756 \nu_{\text{eff}}^3\gamma_L(t)/(\gamma_T^3\sqrt{|A|})$ \cite{devin2025emergence}, where we now include a time-dependent linear growth rate (for details, see Appendix B). An exact solution is obtained by multiplying the equation by $3/2\sqrt{|A|}$ and applying an integrating factor:
\begin{equation} \label{eq:strong_solution_general}
\begin{aligned}
\left|\frac{A(t)}{A_{\text{sat}}}\right|\!\!=\!e^{-\gamma_{d}(t-t_0\!)}\! \!\left[\! \left| \frac{A(t_0)}{A_{\text{sat}}}\right|^{\frac{3}{2}}\!\!\!\!+\!\frac{3 \gamma_d}{2 \gamma_T}\!\int_{t_0}^{t} \! \! \!dt' e^{\frac{3}{2}\gamma_{d}(t'-t_0)}\gamma_L(t')\right]^{\frac{2}{3}}
\end{aligned}
\end{equation}
where $|A_{sat}|^{3/2} \equiv 1.756 \nu_{\text{eff}}^3/ (\gamma_T^2 \gamma_d)$ is the saturation level in the strongly driven limit as $t \rightarrow \infty$ \cite{PetviashviliThesis, BerkBreizman1990a}. To enforce continuity between Phase I and Phase II, $A(t_0)$ defines the amplitude attained at the end of Phase I. In the weakly collisional limit, $A(t_0) \approx 10.2(\gamma_L^2(t_0)/\gamma_T^2)$ \cite{FriedSagdeev1972}; however, when collisions are strong (i.e., $(\gamma_L(t)-\gamma_d)/\nu_{\text{eff}} \ll 1$), the mode may attain a larger amplitude in Phase I \cite{devin2025emergence}. We also define $t_0$ as the time of transition between Phases I and II. We solve Eq. \ref{eq:strong_solution_general} for the form of $\gamma_L(t)$ taken in Eq. \ref{eq:gammaLt}, yielding the following closed-form expression for Phase II evolution:
\begin{equation} 
\begin{aligned}
\label{eq:strong_solution_explicit}
&\left|\frac{A(t)}{A_{sat}}\right|^{\frac{3}{2}} \! \! \! =\!\! \left(\!\left|\frac{A(t_0)}{A_{sat}}\right|^{\frac{3}{2}} \! \! \! -1 \! \right)e^{-3\gamma_{d}(t-t_0)}\! +\!1\!
-\!\frac{3\gamma_d}{2\gamma_T}
\left(\frac{\alpha \gamma_T^3}{\nu_{\text{eff}}^3}\right)^{\frac{3}{2}}\! \! \! \times \\
& \times \! \! \Biggl\{\! \!
\sqrt{\frac{6\pi\gamma_d}{\gamma_T}}
e^{-\frac{3\gamma_d}{2\gamma_T}x^2(t)}\!
\left[
y(t) - y(t_0)\right]\!
- \!
\frac{2}{x(t)}\!
+\!
\frac{
2e^{-\frac{3}{2}\gamma_d(t-t_0)}
}{
x(t_0)
}\! \!
\Biggr\},
\end{aligned}
\end{equation}
where $x(t) \! \equiv \!
\sqrt{\!\gamma_Tt\!+\!\alpha\gamma_T^3/\nu_{\text{eff}}^3}$ and $y(t)\! \equiv\!
\text{erfi}\!\left(\!x(t)\sqrt{3\gamma_d/2\gamma_T}\!\right)$.
\begin{figure}[ht]
\centering
\includegraphics[width=\columnwidth]{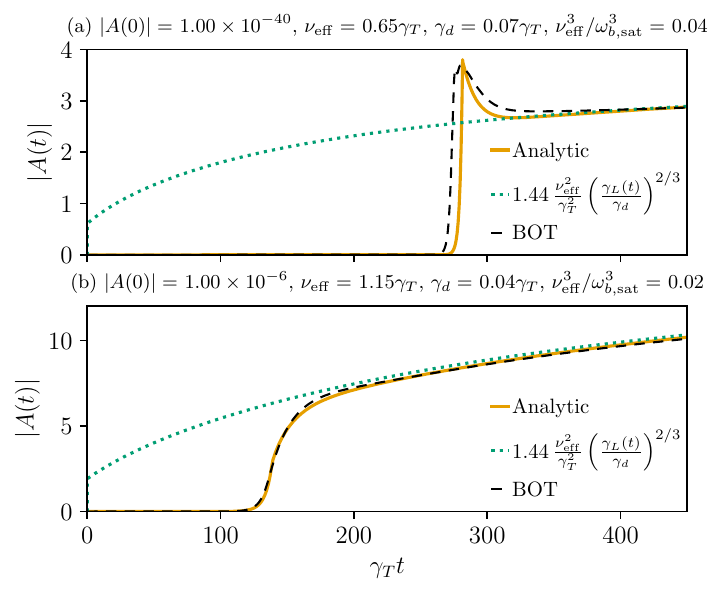}
\caption{Mode amplitude, analytic prediction (Eqs. \ref{eq:strong_solution_phase1} and \ref{eq:strong_solution_explicit}), and time-dependent saturation levels (Eq. \ref{eq:A_far}) for strongly driven modes with (a) $\gamma_d/\gamma_T = 0.07$, $\alpha = 100$ and (b) $\gamma_d/\gamma_T = 0.04$, $\alpha = 1250$. For (a), we take $A(t_0) \approx 10.24(\gamma_L(t_0)/\gamma_T)^2$; for (b), we take $A(t_0) \approx 38(\gamma_L(t_0)/\gamma_T)^2$ \cite{devin2025emergence}.}
\label{fig:driven_amplitude}
\end{figure}
We compare Eqs. \ref{eq:strong_solution_phase1} and \ref{eq:strong_solution_explicit} against BOT runs for small $\gamma_d/\gamma_T$ (Fig. \ref{fig:driven_amplitude}). Both cases satisfy $\omega_b^3/\nu_{\text{eff}}^3 \gg 1$ in the limit $t \rightarrow \infty$. The first case has very small $\omega_b(0)$, which allows the resonant distribution to build up a steep gradient without influence from the mode (which corresponds to the problem in which the distribution is already fully formed). In Fig. \ref{fig:driven_amplitude}(a), the mode grows linearly and significantly overshoots its saturation amplitude before rapidly damping towards the eventual saturation level. The analytic solution described in Eqs. \ref{eq:strong_solution_phase1} and \ref{eq:strong_solution_explicit} matches well with this behavior. The BOT simulated amplitude undergoes rapid growth before the analytic prediction, likely because the extent of mode-induced flattening becomes comparable to $\sqrt{\alpha}$, reducing the accuracy of Eq. \ref{eq:gammaLt}. This limitation of our method also explains the slight discrepancy between $M_L(t)$ and $M_{NL}(t)$ for early times in Fig. $\ref{fig:BOT_run_combined}$(a), and is discussed further in Appendix A. 

\begin{figure}[ht]
\centering
\includegraphics[width=\columnwidth]{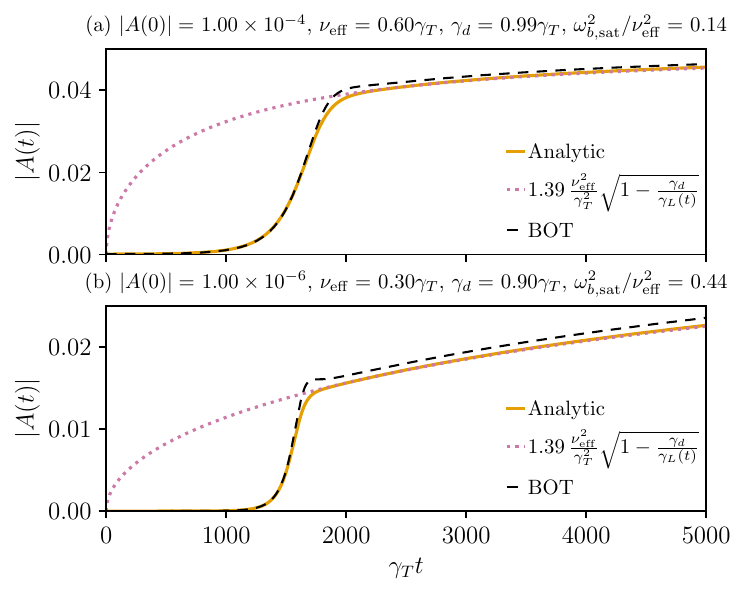}
\caption{Mode amplitude, analytic prediction from Eqs. \ref{eq:marg_solution_explicit} and \ref{eq:marg_kernel_explicit}, and time-dependent saturation levels (Eq. \ref{eq:A_near}) for near-marginal modes with (a) $\gamma_d/\gamma_T = 0.99$, $\alpha=500$ and (b) $\gamma_d/\gamma_T = 0.9$, $\alpha=500$.}
\label{fig:marg_amplitude}
\end{figure}

In contrast, the second case (Fig. \ref{fig:driven_amplitude} (b)) includes a large $\omega_b(0)$ and a slower buildup of the resonant distribution. The mode evolution appears qualitatively similar to the marginally stable case (compare to Fig. \ref{fig:marg_amplitude} for reference), but the analytic solution still matches almost perfectly. Therefore, despite the significant differences between the two BOT calculations, the analytic solution described in Eqs. \ref{eq:strong_solution_phase1} and \ref{eq:strong_solution_explicit} is fairly robust. This further reinforces the importance of $\omega_b/\nu_{\text{eff}}$ as the quantity which determines the evolution and saturation of the mode, independent of the rate of change of the distribution.

\textit{Analytic evolution of marginally unstable modes}\textemdash The other tractable limit is $M_L(t) \ll 1$. As described in \cite{BerkPRL1996,BerkPPR1997, BreizmanPoP1997}, the kinetic equation allows for expansion in $\epsilon \equiv \omega_b^2/\nu_{\text{eff}}^2\ll1$. A generalized version of the cubic equation for a time-dependent linear growth rate can be obtained by following the same steps as in \cite{BerkPRL1996} and \cite{BerkPPR1997}. To third order in $\epsilon$:
\begin{equation} \label{eq:cubic_equation}
\begin{aligned}
&\frac{dA(t)}{dt} \!=\! (\gamma_L(t) \!-\! \gamma_d)A(t) \!-\! \frac{\gamma_T^4}{2}\!\!\int_0^{t/2} \!\!\!\!\!\!dt' t'^2 A(t-t')\int_0^{t-2t'}\!\!\!\!\!\!\!dt'' \times \\
& A(t\!-\!t'\!-\!t'') A^*(t\!-\!2t'\!-\!t'')  \gamma_L(t\!-\!2t'\!-\!t'') e^{ -\nu_{\text{eff}}^3 t'^2\left( \frac{2}{3} t' \!+ t''\right)}.
\end{aligned}
\end{equation}
Following \cite{duarte2018analytical}, the additional approximation $(\gamma_L(t)-\gamma_d)/\nu_{\text{eff}} \ll 1$ allows the time history of the mode amplitude and growth rate to be neglected (see Appendix C for details). We now define a constant $c \equiv \left(3/2\right)^{1/3} \Gamma \left(1/3\right)\gamma_T^4/6 \nu_{\text{eff}}^4$, and after integrating, the cubic equation (Eq. \ref{eq:cubic_equation}) reduces to a time-local Landau-Stuart equation:
\begin{equation} \label{eq:landau_stuart}
\begin{aligned}
\frac{dA(t)}{dt} = &\left(\gamma_L(t) - \gamma_d\right)A(t) - c \gamma_L(t) A(t) |A(t)|^2.
\end{aligned}
\end{equation}
In the limit $t \rightarrow \infty$, we recover the expected saturation amplitude $A_{sat} = \left(1-\gamma_d/\gamma_T\right)^{1/2}c^{-1/2}$  \cite{BerkPPR1997,BreizmanPoP1997}. We analytically solve Eq. \ref{eq:landau_stuart} to explicitly obtain the time evolution of the mode (the details of this computation are contained in Appendix C). using the explicit form of $\gamma_L(t)$ defined in Eq. \ref{eq:gammaLt}:
\begin{equation} \label{eq:marg_solution_explicit}
\begin{aligned}
\!A^{-2}(t) \!=
\!A^{-2}(0) e^{\Phi( t,0)}\!\!+\!2c\gamma_T \!\!\int_0^{t} \!\!\!dt'\!\!
\left[
1\!-\!\left(\!\!\frac{\alpha\gamma_T^3/\nu_{\text{eff}}^3}
{x^2(t')}\!\!\right)^\frac{3}{2}
\right]\!\!
e^{\Phi(t,t')}
\end{aligned}
\end{equation}
where
\begin{equation} \label{eq:marg_kernel_explicit}
\begin{aligned}
\!\!\!\!\!\Phi(t,t') \!\equiv \!
-\!2(\gamma_T\!-\!\gamma_d)(t\!-\!t')\!-\!4\!\left(\!\frac{\alpha \gamma_T^{3}}{\nu_{\text{eff}}^{3}}\!\right)^{\frac{3}{2}}\!\!
\left[\!
\frac{1}{x(t)}\!
-\!
\frac{1}{x(t')}\!
\right]\!.
\end{aligned}
\end{equation}
In the limit of a static distribution (corresponding to $\alpha = 0$), we recover the solution found in \cite{duarte2018analytical}. In the time-dependent case, the behavior is qualitatively similar, as demonstrated in Fig. \ref{fig:marg_amplitude}. However, Eqs. \ref{eq:marg_solution_explicit} and \ref{eq:marg_kernel_explicit} are additionally able to capture the effects of a time-dependent $M_L(t)$. In Fig. \ref{fig:marg_amplitude} (a), the mode is very close to marginal stability, and $\omega_{b,\text{sat}}^2/\nu_{\text{eff}}^2 \ll 1$ is fulfilled. In contrast, Fig. \ref{fig:marg_amplitude}(b) shows that the analytic solution remains fairly accurate even as the mode begins to depart from marginal stability (i.e., as $\omega_{b,\text{sat}}^2/\nu_{\text{eff}}^2$ becomes O(1)).

\begin{figure}[ht]
\centering
\includegraphics[width=\columnwidth]{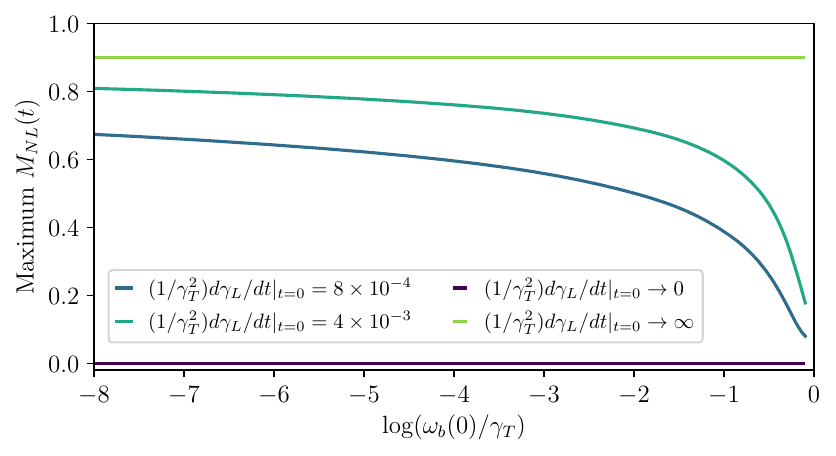}
\caption{Maximum $M_{NL}(t)$ attained during mode evolution, for a range of initial amplitudes and $\gamma_d/\gamma_T = 0.1$. Plotted here are two values of the rate of formation of the distribution $(1/\gamma_T^2)d\gamma_L/dt|_{t=0}$, as well as the limits of infinitely quickly and slowly forming distributions.}
\label{fig:init_amplitude_sweep}
\end{figure}

\textit{Theoretical findings}\textemdash The initial mode amplitude and the rate of distribution formation have been found to be key in determining the character of a kinetic instability in a dynamically forming background. In Fig. \ref{fig:init_amplitude_sweep}, we conduct a systematic scan over these parameters, where the initial mode amplitude is $\omega_b(0)/\gamma_T$ and the rate of distribution formation is defined as $(1/\gamma_T^2)d\gamma_L/dt|_{t=0}$. We then evaluate the maximum $M_{NL}(t)$ attained during mode evolution. High maximum $M_{NL}(t)$ corresponds to strong linear growth, resembling the growth phase using a fully formed distribution (see Fig. \ref{fig:BOT_run_combined}(a)), while cases with low maximum $M_{NL}(t)$ reach a quasi-steady saturation when the underlying EP distribution is still, and thereafter near-marginal (see Fig. \ref{fig:BOT_run_combined}(b)). Fig. \ref{fig:init_amplitude_sweep} shows that the maximum $M_{NL}(t)$ is highest for very small $\omega_b(0)/\gamma_T$, because the resonant distribution is able to develop a steep gradient before the mode grows large enough to substantially reduce its slope. This leads to a strong and extended linear growth phase, similar to the problem where the resonant distribution starts fully formed. However, when $\text{log}(\omega_b(0)/\gamma_T) \gtrsim -1$, the maximum $M_{NL}(t)$ decreases rapidly, and the strong linear growth phase is avoided as the mode is able to reach quasi-steady saturation before the instantaneous linear growth rate significantly exceeds the damping rate. Larger $(1/\gamma_T^2)d\gamma_L/dt|_{t=0}$ also corresponds to a higher maximum $M_{NL}(t)$, as the background gradient forms more quickly.

The long-term behavior and saturation level are independent of the time history of the mode amplitude. The saturation level is determined solely by the value of the linear marginal stability parameter $M_L(t)$ and is unaffected by the nonlinear marginal stability parameter $M_{NL}(t)$. This is because in the presence of sources and sinks, the mode saturation level is determined by the steady-state kinetic equation together with power balance 
(Eqs. \ref{eq:kinetic_eq} and \ref{eq:power_balance}) through an asymptotic equilibrium between replenishment of the resonant distribution free energy and background dissipation. 

We showed that the mode behavior in the case of a dynamically forming distribution can be modeled analytically to first order in the following limits: $\omega_b^3/\nu_{\text{eff}}^3 \gg 1$ for strongly driven modes and $\omega_b^2/\nu_{\text{eff}}^2 \ll 1$  for marginally stable modes, which are the same limits that are typically taken for a fully formed distribution. We observe that near-marginal modes satisfying $M_L(t) \ll 1$ generally satisfy $\omega_b^2/\nu_{\text{eff}}^2 \ll 1$ and that primarily strongly-driven modes with $M_L(t) \approx 1$ satisfy $\omega_b^3/\nu_{\text{eff}}^3 \gg 1$ after the linear growth phase. This means that the strength of the linear drive $M_L(t) \ll 1$ determines the parameter $\omega_b/\nu_{\text{eff}}$ even for time-dependent distributions. However, it is not the case that $M_{NL}(t) \ll 1$ generally implies $\omega_b^2/\nu_{\text{eff}}^2 \ll 1$, as it does for a fully formed resonant distribution. 

\textit{Experimental relevance}\textemdash In a tokamak, the rate of resonant distribution buildup may vary depending on whether the relaxation of the EP distribution is dominated by drag or diffusion. For instance, injected neutral beam ions initially slow down due to dynamical friction on electrons, and may take several tens of milliseconds to travel between injection and a resonance in phase space. However, because drag preserves phase coherence, the "front" of slowing injected particles crosses the resonance on the effective collisional timescale, typically on the order of 0.1-1ms \cite{DuarteAxivPRL}. This would result in quick formation of the resonant distribution, and the mode dynamics would be identical to the case with a fully formed distribution. A near-instantaneous formation of the resonant distribution may also occur in abrupt redistribution events \cite{devin2025emergence}, like Alfvénic evolution following sawteeth relaxations \cite{Kramer2000, kolesnichenko1994thermonuclear} or the thermonuclear instability \cite{kolesnichenko1967}. In contrast, if the EP relaxation is dominated by diffusion, the resonant distribution may form slowly around the resonance if the injection is sufficiently far from the resonance in $\Omega$-space. A kinetic instability may then display the wider range of behaviors reported in this study.

To discuss possible regimes in which the behaviors discussed here might be observed, we contextualize our results based on reported experimental measurements of EP driven instabilities. In Fig. \ref{fig:init_amplitude_sweep}, the intermediate values of $(1/\gamma_T^2)d\gamma_L/dt|_{t=0}$ correspond to distributions which build up on a timescale of 250ms and 500ms, comparable to the slowing-down time of a low density tokamak \cite{Gaffey1976JPP}, but much slower than the effective collisional time within the mode resonance \cite{SuOberman1968}. In such a slowly-forming resonant distribution, Fig. \ref{fig:init_amplitude_sweep} implies that modes with $\gamma_d/\gamma_T = 0.1$ remain close to their instantaneous saturation for the duration of their evolution when $\text{log}(\omega_b(0)/\gamma_L) \gtrsim -1$. On the other hand, if the distribution were to build up on the effective collisional timescale of approximately $0.1 \text{ms}$, then $(1/\gamma_T^2)d\gamma_L/dt|_{t=0} \approx 0.5$, several orders of magnitude higher than the other cases shown in Fig. \ref{fig:init_amplitude_sweep}. This case is represented by the $(1/\gamma_T^2)d\gamma_L/dt|_{t=0} \rightarrow \infty$ limit, where the dynamics are identical to those in a fully formed distribution. Finally, if the distribution forms on an infinitely long timescale ($(1/\gamma_T^2)d\gamma_L/dt|_{t=0} \rightarrow 0$), modes will always remain near their instantaneous nonlinear threshold.

In previous analyses of toroidal Alfvén eigenmodes excited by ion cyclotron resonant heating on TFTR \cite{WongBerkSaturationinTFTRPoP1997}, it was estimated that $\gamma_L \approx 1.6\times10^{4} s^{-1},~\omega_{b,\text{sat}} \approx 8.5 \times 10^3 s^{-1},~\nu_{\text{eff}} \approx 8.2\times 10^3 s^{-1}$, and $\omega_b(0) \approx 0.3 \omega_{b,\text{sat}} \approx 2.7\times 10^3 s^{-1}$. Note that NSTX \cite{WhiteNSTX2020} and DIII-D \cite{spong2021nonlinear} also found mode saturation amplitudes to be about one order of magnitude larger than the background fluctuation level. From this, we estimate $\text{log}(\omega_b(0)/\gamma_L) \approx -0.7$, which in the context of Fig. \ref{fig:init_amplitude_sweep}, implies that modes may indeed remain close to their instantaneous nonlinear threshold for a slowly forming distribution. 

While this study has only considered local dynamics near a resonance, future work will investigate kinetic instabilities where the global formation of the resonant distribution is treated realistically, possibly using an orbit-tracing code. Such work will also consider the effects of a realistic geometry and mode structure, and enable direct comparisons of our theory to experiment.

\textit{Acknowledgments}\textemdash We thank M. K. Lilley for developing and making the BOT code openly available. This manuscript is based upon work supported by the US Department of Energy, Office of Science, Office of Fusion Energy Sciences, and has been authored by Princeton University under Contract DE-AC02-09CH11466 with the US Department of Energy. The work was supported by the DOE Early Career Research Program, project \textit{Phase-Space Engineering of Supra-Thermal Particle Distribution for Optimizing Burning Plasma Scenarios}. The publisher, by accepting the article for publication, acknowledges that the United States Government retains a non-exclusive, paid-up, irrevocable, world-wide license to publish or reproduce the published form of this manuscript, or allow others to do so, for United States Government purposes.

\bibliographystyle{apsrev4-2} 

%

\appendix
\section{Modification of the BOT code for a time-dependent distribution}
In its standard implementation, the BOT code \cite{Lilley2010} assumes that the normalized distribution function $G \equiv (2\pi e^2\omega/mk\gamma_T^2)f$ is linear around the resonance with slope $1/\pi$ in $\Omega$-space. It numerically solves for the Fourier harmonics in $\Omega$-space after integrating analytically once over characteristics. Without this uniform-slope background, the integration over characteristics cannot be performed. BOT also does not include any global structure for the underlying EP distribution; therefore, we can only change the local structure of the distribution near the resonance. To allow a linear growth rate $\gamma_L(t)$ that starts at zero and grows towards the target growth rate $\gamma_L(t\rightarrow\infty)\equiv\gamma_T$, we modify the initial condition so that the EP distribution starts as a plateau centered on the resonance:
\begin{equation}
G_T +g_0(\Omega, t=0) = \frac{\omega+\Omega}{\pi}-\frac{\Omega}{\pi}\exp(-\Omega^2/4\alpha)~.
\end{equation}
Here, $\alpha$ is the parameter that determines the width of the plateau. Note that this initial condition asymptotically approaches the target distribution $(\omega+\Omega)/\pi$ as $\Omega \rightarrow \infty$, which remains true as the distribution evolves. In order to still consider the distribution linear around the resonance, we must choose $\alpha$ such that the characteristic width of the resonance at saturation is much smaller than $\sqrt\alpha$. For strongly driven modes in particular, the resonance can widen enough to access parts of the distribution with steeper gradients. If $\sqrt{\alpha}$ is chosen to be small, which is necessary to access strongly-driven regimes on reasonable timescales, the linear growth rate will be larger than that calculated in Eq. \ref{eq:gammaLt}. This especially affects the accuracy of Eq. \ref{eq:strong_solution_phase1}, which will be delayed relative to the BOT calculated mode amplitude. 

We allow collisions to relax the distribution towards its assumed linear form, which results in a target growth rate $\gamma_T$. To quantify the linear growth rate as a function of time for diffusive collisions, we analytically solve the kinetic equation in the absence of modes:
\begin{equation} 
\frac{\partial G}{\partial t} = \frac{\nu_{\text{eff}}^3}{\gamma_T^2}\frac{\partial^2}{\partial \Omega^2} \left(G-G_T\right)~.
\end{equation}
Here, $\nu_{\text{eff}}$ is the effective scattering rate, $\Omega$ is the normalized velocity, and $G = G_T + g_0$ is the total background distribution. The solution is:
\begin{equation}
G(\Omega,t) = \frac{\omega}{\pi}+\frac{\Omega}{\pi} \Biggl\{1-\frac{\text{exp}\left[ -\Omega^2/\left(4\nu_{\text{eff}}^3t/{\gamma_T^2} +4\alpha\right)\right]}{ \left(\nu_{\text{eff}}^3t/{(\alpha\gamma_T^2)} +1\right)^{3/2}} \Biggr\}.
\end{equation}
This solution recovers the initial condition at $t = 0$ and is forced towards the target distribution as $t \rightarrow \infty$. We also calculate the linear growth rate at the resonance $\Omega = 0$:
\begin{equation}
\gamma_L(t) =  \gamma_T \left[1-\frac{\alpha^{3/2}}{\left(\nu_{\text{eff}}^3t/{\gamma_T^2} +\alpha\right)^{3/2}}\right]~.
\end{equation}
While our study uses this particular form of $\gamma_L(t)$ due to the constraints imposed by the BOT code, we do not expect other forms of $\gamma_L(t)$ to qualitatively change the physics reported here. At $t = 0$, the characteristic rate of change of $\gamma_L$ is
\begin{equation}
\frac{1}{\gamma_T^2}\frac{d\gamma_L}{dt} \sim \frac{\nu_{\text{eff}}^3}{\alpha \gamma_T^3}.
\end{equation}
Thus, the rate at which the linear growth rate changes is determined by the effective scattering rate $\nu_{\text{eff}}$ as well as the parameter $\alpha$. In a plasma, the effective scattering controls the linear growth rate through relaxation of the distribution in cases where the EPs undergo diffusion into the resonance after they are injected, for example in ICRH-heated plasmas. In our model, $\alpha$ determines the inhomogeneity of the background distribution and therefore influences its diffusive relaxation rate. 

The above modifications to BOT can also be made for a Krook collision operator, $C[f]=\nu_K \left(F_T-f\right)$, though in the Krook case the distribution relaxes on exactly the effective collisional timescale. This approach may be better suited for cases where the distribution changes on the same timescale as mode growth. For instance, when NB ions undergo a slowing-down process they cross the resonance on the effective collisional timescale, which may be comparable to the time for mode growth. For a mode with a steady saturation level, $\nu_K$ is large enough that the distribution relaxes almost instantaneously relative to the mode growth. For this reason, we primarily focus on diffusive collisions in this work.

\section{Amplitude equation for strongly driven instabilities in the time-dependent case}

To extend the analytical treatment of strongly-driven modes of Ref. \cite{devin2025emergence}, we assume an unperturbed equilibrium of a constant slope scaled by the linear growth rate given in Eq. \ref{eq:gammaLt}:

\begin{equation}
    G_0(t,\Omega) \approx \frac{\omega}{\pi} + \frac{\Omega}{\pi} \left(1-\frac{\alpha^{3/2}}{\left(\nu_{\text{eff}}^3 t/\gamma_T^2+\alpha\right)^{3/2}}\right) = \frac{\omega}{\pi} +\frac{\Omega}{\pi} \frac{\gamma_L(t)}{\gamma_T},
\end{equation} 
where $\Omega = k(v-\omega/k)/\gamma_T$.  
Since $\gamma_L(t)$ does not depend on $\Omega$, it will simply appear as a multiplicative factor in the leading order distribution function.  For trapped and passing particles respectively, this is given by

\begin{equation}\label{eq: strongly driven f0}
\frac{\partial g_0^{\pm}}{\partial E} =
\begin{cases}
\displaystyle
0, ~~~~~ -A\leq E\leq A\\
\displaystyle
\frac{\pm 2\gamma_L(t)/\gamma_T} {\int_{-3\pi/2}^{\pi/2}\mathrm{d}z \sqrt{2(E+|A|\sin z)}}, & E<A
\end{cases}
\end{equation}
where $E = \Omega^2/2 -|A|\sin z$, and $z = kx-\omega t +\phi$. The next order is found to be 

\begin{equation} \label{eq: strongly driven f1}
    \frac{\partial g_1^\pm}{\partial z} = \pm \frac{\nu_{\text{eff}}^3}{\gamma_T^3}  \frac{\partial }{\partial E} \sqrt{2(E+|A|\sin z)} \frac{\partial g_0^\pm }{\partial E}
\end{equation}
 which will therefore also be proportional to $\gamma_L(t)$.
The amplitude equation is given by 

\begin{equation}
\frac{d|A|}{dt}+\gamma_{d}|A|=\frac{\gamma_T\sqrt{2|A|}}{\pi}\int_{-\pi}^{\pi}\mathrm{d}z\int_{-\sin z}^{\infty}\mathrm{d}y\, w\frac{\partial (g^++g^-)}{\partial z}
\end{equation}
where $w = \sqrt{y+\sin z}$, and $y = E/|A|$. Substituting Eqs. \ref{eq: strongly driven f0} and \ref{eq: strongly driven f1} in for $f^\pm$, the integrals are of identical form to Refs. \cite{PetviashviliThesis,devin2025emergence} and the final amplitude equation is given by


\begin{equation}
\frac{d|A|}{dt}+\gamma_{d}|A|=1.756 \frac{\nu_{\text{eff}}^3 \gamma_L(t)}{\gamma_T^3\sqrt{|A|}}.
\end{equation}
\section{Reduction of the cubic equation to a Landau-Stuart form for near-marginal instabilities}

Consider the time-delayed cubic equation (Eq. \ref{eq:cubic_equation}):
\begin{equation}
\begin{aligned}
&\frac{d\omega_b^2}{dt} = (\gamma_L(t) - \gamma_d)\omega_b^2 - \frac{1}{2}\int_0^{t/2} dt' t'^2 \int_0^{t-2t'}dt'' \times \\
&\times\Biggl\{ \omega_b^2(t-t')\omega_b^2(t-t'-t'') \omega_b^{2*}(t-2t'-t'')\times \\
&\times \gamma_L(t-2t'-t'') \text{exp}\left[ -\nu_{\text{eff}}^3 t'^2\left( \frac{2}{3} t' + t''\right)\right]\Biggr\}~.
\end{aligned}
\end{equation}
We first address the inner integral
\begin{equation}
\begin{aligned}
&I_1 
\equiv  \int_0^{t-2t'}dt'' \Biggl\{ \omega_b^2(t-t'-t'') \omega_b^{2*}(t-2t'-t'') \times \\ 
&\times \gamma_L(t-2t'-t'') \text{exp}\left[ -\nu_{\text{eff}}^3 t'^2t''\right]\Biggr\}~,
\end{aligned}
\end{equation}
and integrate it by parts:
\begin{equation}
\begin{aligned}
&I_1 = \frac{1}{\nu_\text{eff}^3 t'^2}\Biggl\{\omega_b^2(t-t'-t'') \omega_b^{2*}(t-2t'-t'') \\
&\times \gamma_L(t-2t'-t'')\text{exp} 
\left[ -\nu_{\text{eff}}^3 t'^2t''\right]\Biggr\}_{t-2t'}^{0} + I_2~,
\end{aligned}
\end{equation}
where 
\begin{equation}
\begin{aligned}
&I_2 = \frac{1}{\nu_\text{eff}^3 t'^2}\int_0^{t-2t'}dt'' \frac{d}{dt''}\Biggl\{ \omega_b^2(t-t'-t'') \omega_b^{2*}(t-2t'-t'') \\
&\times \gamma_L(t-2t'-t'')\Biggr\}\text{exp}\left[ -\nu_{\text{eff}}^3 t'^2t''\right]~.
\end{aligned}
\end{equation}
Focusing on $I_2$ , we note that:
\begin{equation}
\frac{d}{dt''}\text{log}(\omega_b^2(t-t'-t''))\lesssim \text{max}\left[ \gamma_L(t) - \gamma_d\right],
\end{equation}
\begin{equation}
\frac{d}{dt''}\text{log}(\omega_b^{2*}(t-2t'-t'') )\lesssim \text{max}\left[ \gamma_L(t) - \gamma_d\right],
\end{equation}
\begin{equation}
\frac{d}{dt''}\text{log}(\gamma_L(t-2t'-t'')) \lesssim \text{max}\left[\frac{d}{dt}\text{log}(\gamma_L(t))\right]~,
\end{equation}
and therefore,
\begin{equation}
I_2 \lesssim \left( \!\frac{2\text{max}[\gamma_L(t) - \gamma_d]}{\nu_{\text{eff}}} \! + \!\frac{\text{max}\left[d\text{log}(\gamma_L(t))/dt\right]}{\nu_{\text{eff}}} \right)\!\!\frac{1}{\nu_\text{eff}^2 t'^2} I_1~.
\end{equation}
The reduction to time-local form requires
\begin{equation}
\frac{\text{max}\left[ \gamma_L(t) - \gamma_d\right]}{\nu_{\text{eff}}} \ll 1~, \frac{\text{max}\left[d\text{log}(\gamma_L(t))/dt\right]}{\nu_\text{eff}} \ll 1~.
\end{equation}
The first requirement is the same as used in previous work on fully formed distributions \cite{duarte2018analytical,duarte2019collisional}, while the second condition is new. The slow formation rate of the distribution compared to $\nu_{\text{eff}}$ is satisfied for the case explored in this work. On the other hand, a Krook collision operator ($C[f] = \nu_\text{K}(F_T - f)$) would result in $d\text{log}(\gamma_L(t))/dt \sim \nu_{\text{K}}$, and this condition would be invalid. With these reductions, we find that $I_1 \leq B \leq (1-\epsilon/(\nu_{\text{eff}}^2 t'^2))I_1$, where $\epsilon$ is a small parameter determined by (Eq. C7) and $B$ is the boundary term in (Eq. C3). Now, returning to the full cubic equation (Eq. C1) and focusing only on terms that do not depend on $\omega_b$ or $\gamma_L$, we see that:
\begin{equation}
\begin{aligned}
&t'^2\left(1-\frac{\epsilon}{\nu_{\text{eff}}^2 t'^2}\right)\text{exp}\left[ -\nu_{\text{eff}}^3 t'^2\left( \frac{2}{3} t' + t''\right)\right]  \equiv \text{exp}[{-\Phi}]\\
& = \text{exp}\left[ -\nu_{\text{eff}}^3 t'^2\left( \frac{2}{3} t' + t''\right) + \text{ln}\left( t'^2 - \frac{\epsilon}{\nu_{\text{eff}}^2}\right)\right]
\end{aligned}
\end{equation}
The structure of the kernel is therefore unchanged when $\epsilon$ is allowed to be finite, since its peak is still located near ($t'' = 0, t' = 1/\nu_{\text{eff}}$) to leading order in $\epsilon$, and in the $t'$ direction, the exponential decay still occurs on the scale of $1/\nu_{\text{eff}}$. To see this rigorously, the exponential term may be treated with Laplace's method in $t'$, where the argument of the exponential is expanded around its minimum before integration:
\begin{equation}
\Phi \approx \Phi(t'_{min}, t'') + \frac{1}{2} (t' -t'_{min})^2\partial^2_{t'}\Phi|_{t_{min}'} + ...
\end{equation}
Using (Eq. C8), it is possible to solve for the quantities $t'_{min},\Phi(t'_{min}, t''), \partial^2_{t'}\Phi|_{t_{min}'}$ perturbatively in $\epsilon$. This implies that $(1-\epsilon/(\nu_{\text{eff}}^2 t'^2))I_1 \rightarrow I_1$ as $\epsilon \rightarrow 0$, in the sense that the leading-order behavior will be the same when integrated over $t'$. Therefore, $I_1 \approx B$ for small $\epsilon$:

\begin{equation}
\begin{aligned}
&I_1 \approx \frac{1}{\nu_\text{eff}^3 t'^2}\Biggl\{\omega_b^2(t-t') \omega_b^{2*}(t-2t')\gamma_L(t-2t') \\
& -\omega_b^2(t') \omega_b^{2*}(0)\gamma_L(0)\text{exp} 
\left[-\nu_{\text{eff}}^3 t'^2(t-2t')\right]\Biggr\}~.
\end{aligned}
\end{equation}
The second boundary term is exponentially suppressed for small values of $t'$, so the cubic equation reduces to
\begin{equation}
\begin{aligned}
&\frac{d\omega_b^2}{dt} = (\gamma_L(t) - \gamma_d)\omega_b^2 - \frac{1}{2\nu_{\text{eff}}^3}\int_0^{t/2} dt' \times \\
& \omega_b^4(t-t') \omega_b^{2*}(t-2t') \gamma_L(t-2t') \text{exp}\left[ -\frac{2}{3}\nu_{\text{eff}}^3 t'^3\right]~.
\end{aligned}
\end{equation}
Using the same justification as for the previous integral, we note that the exponential term is strongly peaked around $t' = 0$ and decays much faster than the net growth rate of $\omega_b$, allowing us to neglect the time history in $t'$. Then, 
\begin{equation}
\int_0^\infty dt' \text{exp}\left[ -\frac{2}{3}\nu_{\text{eff}}^3 t'^3\right] = \frac{1}{3\nu_{\text{eff}}} \left( \frac{3}{2} \right)^{1/3} \Gamma\left(\frac{1}{3} \right)~.
\end{equation}
Using this result, and additionally defining $A \equiv \omega_b^2/\gamma_T^2$, we recover Eq. \ref{eq:landau_stuart}:
\begin{equation}
\begin{aligned}
\frac{dA(t)}{dt} = &\left(\gamma_L(t) - \gamma_d\right)A(t) - c \gamma_L(t) A(t) |A(t)|^2~,
\end{aligned}
\end{equation}
where $c \equiv \left(3/2\right)^{1/3} \Gamma \left(1/3\right)\gamma_T^4/6 \nu_{\text{eff}}^4$. The general solution to this equation is obtained by following \cite{duarte2018analytical}. We divide by $A^{-2}$ to obtain
\begin{equation}
\frac{d}{dt} (A^{-2}) + 2\left( \gamma_L(t) - \gamma_d\right) A^{-2} = 2c\gamma_L(t)~.
\end{equation}
This equation can then be solved with an integrating factor. We then find that
\begin{equation}
A^{-2}(t) = 
A^{-2}(0) e^{\Phi(t,0)}
+ 2c \int_0^{t} dt'
\gamma_L(t')
e^{\Phi(t,t')}~,
\end{equation}
where we define
\begin{equation}
\Phi(\tau,\tau') \equiv 2\int_{\tau}^{\tau'}d\tau'' (\gamma_L(\tau'')-\gamma_d)~.
\end{equation}

\end{document}